\documentclass[twocolumn,showpacs,10pt]{revtex4-1}%
\usepackage{amsmath,amssymb,graphicx}
\usepackage{amsmath}
\usepackage{amsfonts}
\usepackage{CJK}
\usepackage{amssymb}
\usepackage{color}
\usepackage{graphicx}%
\providecommand{\U}[1]{\protect\rule{.1in}{.1in}}
\begin{document}

\title{Quantum image rain removal: second-order photon number fluctuation
correlations in the time domain}
\author{Yuge Li, Yunjie Xia, Deyang Duan}
\email{duandy2015@qfnu.edu.cn}
\affiliation{School of Physics and Physical Engineering, Qufu Normal University, Qufu 273165, China}
\begin{abstract}
Falling raindrops are usually considered purely negative factors for
traditional optical imaging because they generate not only rain streaks but
also rain fog, resulting in a decrease in the visual quality of images.
However, this work demonstrates that the image degradation caused by falling
raindrops can be eliminated by the raindrops themselves. The temporal
second-order correlation properties of the photon number fluctuation
introduced by falling raindrops has a remarkable attribute: the rain streak
photons and rain fog photons result in the absence of a stable second-order
photon number correlation, while this stable correlation exists for photons
that do not interact with raindrops. This fundamental difference indicates
that the noise caused by falling raindrops can be eliminated by measuring the
second-order photon number fluctuation correlation in the time domain. The
simulation and experimental results demonstrate that the rain removal effect
of this method is even better than that of deep learning methods when the
integration time of each measurement event is short. This high-efficient quantum
rain removal method can be used independently or integrated into deep learning algorithms to provide
front-end processing and high-quality materials for deep learning.

\end{abstract}
\maketitle

Rain is a common weather phenomenon, but its presence significantly reduces
the visual quality of objects in captured images. Consequently, it degrades
the performance of many computer vision tasks, e.g., object recognition,
surveillance and autonomous driving. Hence, rain removal has long been a
fundamental problem in computer vision. For image rain removal, early methods
focused on removing rain streaks by using image priors [1-5]. However, these
image priors are not always reliable. In recent years, various methods based
on deep learning (convolutional neural networks, generative adversarial
networks) have been proposed to overcome the drawbacks of using image priors
[6-14]. Deep learning algorithms resort to training on synthetic data.
However, model training requires large quantities of paired hazy/clean images,
and it is almost impossible to obtain these image pairs from the real world.
Nevertheless, image rain removal is almost monopolized by deep learning
technology [1-15].

Image rain removal is a challenging task since noise caused by rain spatially
varies across rainy images [15]. Obviously, the scene visibility spatially
varies in the image space, since objects closer to the camera are affected
mainly by rain streaks, while objects far from the camera are affected more
heavily by fog (Fig. 1a). This phenomenon is depicted by the model proposed by
Garg and Nayar [16]. In classical optics, an image is produced by the
measurement of the two-dimensional intensity distribution of the light field.
From the perspective of coherence, directly measuring the intensity of light
can be used to obtain the first-order correlation of the light field [17,18].
Consequently, the rain streaks and rain fog introduced by raindrops are
usually considered a purely negative factor [16]. However, studies on the
second-order correlation of photons interacting with rain medium have not been
reported thus far. In this article, we first investigate the temporal
statistical correlation properties of photons interacting with falling
raindrops. The temporal second-order photon number correlation difference
between rain streaks photons, rain fog photons, and photons that do not
interact with raindrops is used to eliminate rain streaks and rain fog. Thus,
a clean image is obtained through the quantum properties of light.
\begin{figure}[ptbh]
\centering
\fbox{\includegraphics[width=0.95\linewidth]{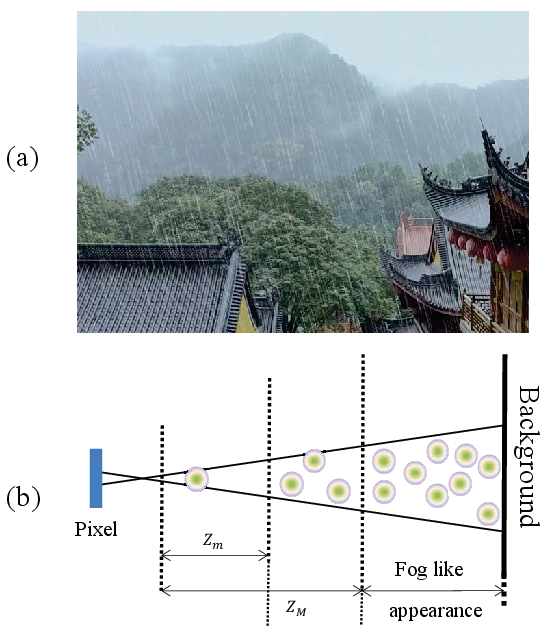}}\caption{(a) An example
of a real photo, in which the scene visibility variation with depth and the
presence of rain streaks and fog are demonstrated; and (b) the change in the
noise produced by falling raindrops as the drop's distance $z$ from the camera
changes. The noise is manifested as rain streaks when $z<z_{M}$. While for
raindrops far from the camera ($z>z_{M}$), the noise manifests as rain fog.}%
\label{fig:false-color}%
\end{figure}

Different from the fog medium [19-24], the degradation of images caused by
rain is more complex. Falling raindrops close to the camera produce rain
streaks, while raindrops far away from the camera cause a fog-like blur [16].
See Figure 1(a) for a real photo example; see Figure 1(b) for the change in
the noise produced by raindrops as a function of the drop's distance $z$ from
the camera. For the fog-like blur, the interaction between raindrops and light
can be described using the McCartney scattering model [25]. Rain streaks are
caused by raindrops occurring in multiple pixels within an integral time.
Consequently, the light received by one pixel described in traditional optics
can be represented as[7-9,11-16]%
\begin{equation}
I\left(  z,t\right)  =\mu\left(  z,t\right)  B\left(  z\right)  +\left(
1-\mu\left(  z,t\right)  \right)  L\left(  z\right)  +R\left(  z\right)  ,
\end{equation}
where $I\left(  z,t\right)  $ represents the light field measured by the
camera, $B\left(  z\right)  $ represents a clear image, $L\left(  z\right)  $
represents the radiance caused by raindrops, $R\left(  z\right)  $ represents
the rain streaks, and $\mu\left(  z,t\right)  $ and $z$ represent the
scattering function and longitudinal coordinate, respectively.

To investigate the statistical properties of photons interacting with falling
raindrops, we adopt the granularity interpretation of radiation instead of the
continuum interpretation of classical optics. This leads to a statistical view
of light: point sources of radiation emit photons or subfields randomly in all
possible directions. The radiation measured at coordinate $(z,t)$ is the
result of a superposition among many random subfields, $\sum_{m=1}^{\infty
}E_{m}\left(  z,t\right)  $, each emitted from a point source. The measured
intensity is proportional to the number of photons. Thus, Eq. 1 is rewritten
as%
\begin{align}
I\left(  \rho_{^{\prime}},z,\lambda\right)   &  =\int_{object}\int_{0}^{T}%
\sum_{m=1}^{N}E_{m}\left(  \rho,t\right)  \mu\left(  z,t\right)  d\rho
dt\nonumber\\
&  +\int_{object}\int_{0}^{T}\sum_{m=1}^{N}E_{m}^{^{\prime}}\left(
\rho,t\right)  \left(  1-\mu\left(  z,t\right)  \right)  d\rho dt\\
&  +\int_{0}^{T}\tau E_{r}\left(  z,t\right)  dt,\nonumber
\end{align}
where $E_{r}$ is the time-averaged irradiance caused by raindrops [16]. The
time $0<\tau<1.18$ ms that a drop projects onto a pixel is far less than the
integral $T\approx30$ ms of a video camera [16]. An object is assumed to be
self-luminous and each point on the object surface is assumed to be an
independent point subsource of radiation, where $\rho$ is the transverse
coordinate of the point source. The photons emitted by the point source
$(\rho)$ pass through the rain medium and are focused onto an image plane
$(\rho^{^{\prime}})$ by the Gaussian thin lens equation $1/s_{0}+1/s_{i}=1/f$,
where $s_{o}$ is the distance between the object and the imaging lens, $s_{i}$
the distance between the imaging lens and the image plane, and $f$ is the
focal length of the imaging lens. Moreover, we assume that (i) $N$ is
proportional to the integral time $T$, and (ii) the number of photons emitted
per unit time on the surface of the object is statistically constant. Thus,
Eq. 2 is rewritten as%
\begin{align}
I\left(  z,T,\tau\right)   &  =\sum_{m=1}^{N}E_{m}\left(  \rho,t\right)
\int_{0}^{T}\mu\left(  z,t\right)  dt\nonumber\\
&  +\sum_{m=1}^{N}E_{m}\left(  \rho,t\right)  \int_{0}^{T}\left(  1-\mu\left(
z,t\right)  \right)  dt\nonumber\\
&  +\tau\sum_{m=1}^{N^{^{\prime}}}E_{m}^{^{\prime}}\left(  z,t\right)
\nonumber\\
&  =S\left(  z,T\right)  +F\left(  z,T\right)  +D\left(  z,\tau\right)  ,
\end{align}
where $S\left(  z,T\right)  $, $F\left(  z,T\right)  $, $D\left(
z,\tau\right)  $ represent the statistical averages of the photon numbers in a
measurement event of the photons that do not interact with raindrops, rain fog
photons, and rain streaks photons, respectively. Equation 3 shows that the
light detected by one pixel exhibits significant photon number fluctuations in
different measurement events. Thus, there are photon number fluctuations on
the entire image plane for different measurement events along the time axis.
\begin{figure}[ptbh]
\centering
\fbox{\includegraphics[width=1\linewidth]{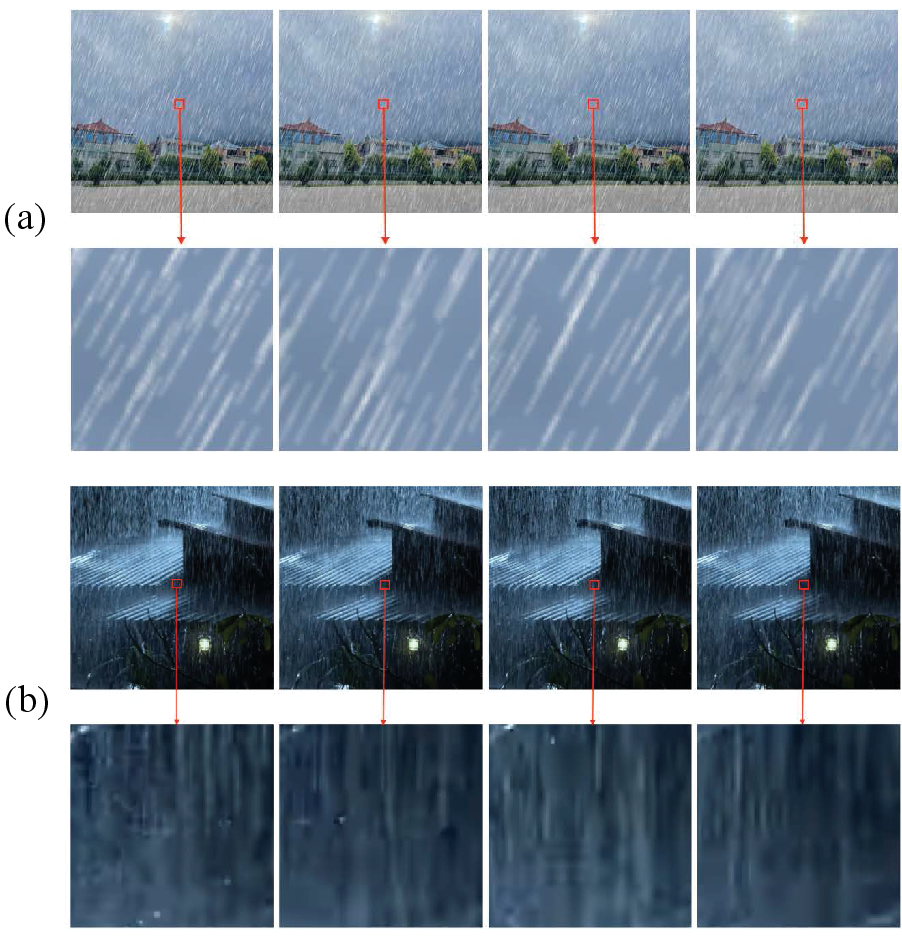}}\caption{The photon number
fluctuation phenomenon on a tiny pixel block for different measurement events
along the time axis. (a) the simulation results and (b) the real rain
results.}%
\label{fig:false-color}%
\end{figure}

A set of simulation and experimental results are used to demonstrate the above
theory. Figure 2 shows that photon number fluctuations do indeed occur on a
tiny pixel block. To measure the photon number fluctuations on the image
plane, we use the peak signal-to-noise ratio (PSNR) index to measure the pixel
block (Fig. 3). For short integration times, the photon number variation due
to rain is high. For long integration times, the intensity at the same pixel
shows low variance over time. This random phenomenon is introduced by a
time-variant rain medium. \begin{figure}[ptbh]
\centering
\fbox{\includegraphics[width=1\linewidth]{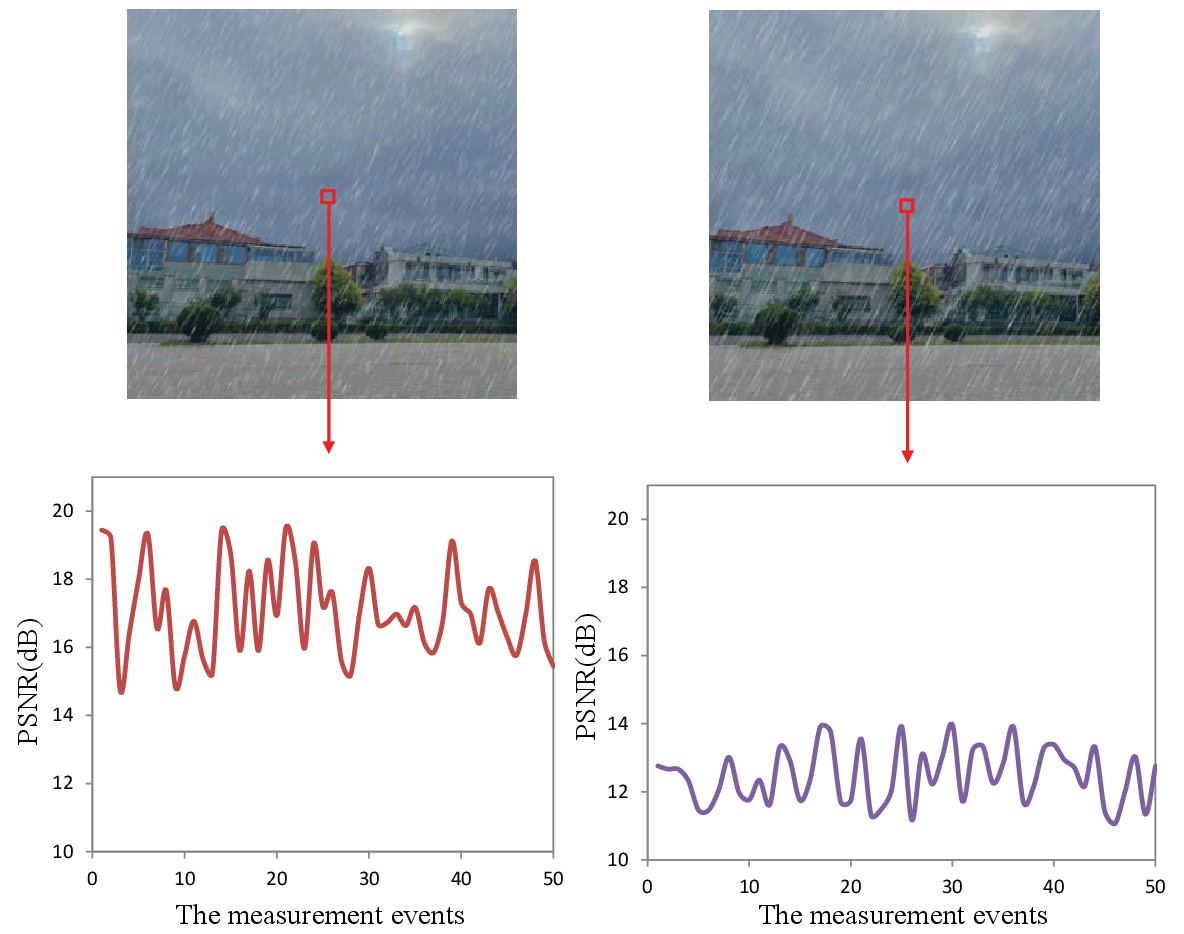}}\caption{Dynamic weather
and photon number fluctuation. The integration times for (a) and (b) are 20 ms
and 50 ms, respectively. }%
\label{fig:false-color}%
\end{figure}

The second-order photon number correlation of any two consecutive measurement
events along the time axis is expressed as [17, 26]%
\begin{align}
&  \left\langle G\left(  z,T_{1},T_{2}\right)  \right\rangle \nonumber\\
&  =\left\langle S_{1}\left(  z,T_{1}\right)  S_{2}\left(  z,T_{2}\right)
\right\rangle +\left\langle F_{1}\left(  z,T_{1}\right)  F_{2}\left(
z,T_{2}\right)  \right\rangle \\
&  +\left\langle D_{1}\left(  z,T_{1}\right)  D_{2}\left(  z,T_{2}\right)
\right\rangle .\nonumber
\end{align}
For the photons that do not interact with raindrops, we have%
\begin{align}
&  \left\langle S_{1}\left(  z,T_{1}\right)  S_{2}\left(  z,T_{2}\right)
\right\rangle \nonumber\\
&  =\left\langle \sum_{m=1}^{N}E\left(  \rho,t_{1}\right)  \sum_{m=1}%
^{N}E\left(  \rho,t_{2}\right)  \right\rangle \nonumber\\
&  \times\left\langle \int_{0}^{T}\mu\left(  z,t_{1}\right)  dt\int_{0}^{T}%
\mu\left(  z,t_{2}\right)  dt\right\rangle \\
&  \approx\left\vert G^{\left(  1\right)  }\left(  t_{1},t_{2}\right)
\right\vert ^{2}\left\langle \mu\left(  T_{1}\right)  \mu\left(  T_{2}\right)
\right\rangle .\nonumber
\end{align}
where $G^{\left(  1\right)  }$ is usually defined as the first-order coherence
function [17,18], $\int_{0}^{T}\mu\left(  z,t\right)  dt\approx\mu\left(
z,T\right)  $ when $T\sim\tau$. Similarly, the second-order photon number
correlations of rain fog photons and rain streak photons can be obtained.
Thus, Eq. 4 is rewritten as%
\begin{align}
\left\langle G\left(  z,T_{1},T_{2}\right)  \right\rangle  &  \approx
\left\vert G^{\left(  1\right)  }\left(  t_{1},t_{2}\right)  \right\vert
^{2}\left\langle \mu\left(  T_{1}\right)  \mu\left(  T_{2}\right)
\right\rangle \nonumber\\
&  +\left\vert G^{^{\prime}\left(  1\right)  }\left(  t_{1},t_{2}\right)
\right\vert ^{2}\left(  1-\left\langle \mu\left(  T_{1}\right)  \mu\left(
T_{2}\right)  \right\rangle \right) \\
&  +\left\vert G^{^{^{\prime\prime}}\left(  1\right)  }\left(  t_{1}%
,t_{2}\right)  \right\vert ^{2}.\nonumber
\end{align}

Considering the coherence time $\Delta t$ of the light field, previous works
showed that when the coherence time $\Delta t>\Delta T$, where $\Delta T$
represents the time interval between two measurement events, $G^{\left(
1\right)  }\neq0$, $G^{^{\prime}\left(  1\right)  }\neq0$, and $G^{^{^{\prime
\prime}}\left(  1\right)  }\neq0$ [27-29]. Thus, a clean image is not
obtained. If $\Delta t<\Delta T$, $G^{\left(  1\right)  }=0$, $G^{^{\prime
}\left(  1\right)  }=0$, and $G^{^{^{\prime\prime}}\left(  1\right)  }=0$.
Thus, no image is obtained. Considering the accumulation of photon numbers
over a period [29], if $\Delta t<\Delta T$, $G^{\left(  1\right)  }\neq0$
because the photons that do not interact with raindrops follow a
point-to-point relationship between the object plane and the image plane. Due
to the refraction caused by falling raindrops [16], the point-to-point
relationship of rain fog photons is disrupted. Consequently, $G^{^{\prime
}\left(  1\right)  }=0$. The randomness of rain results in a completely random
distribution of rain streak photons on the image plane. Thus, $G^{^{^{\prime
\prime}}\left(  1\right)  }=0$. In practice, $G^{^{\prime}\left(  1\right)
}\rightarrow0$ and $G^{^{^{\prime\prime}}\left(  1\right)  }\rightarrow0$
because of the photon number accumulation of the CCD detector. Moreover,
$\left\langle \mu\left(  T_{1}\right)  \mu\left(  T_{2}\right)  \right\rangle
\propto1/T$, $0<\left\langle \mu\left(  T_{1}\right)  \mu\left(  T_{2}\right)
\right\rangle <1$. Consequently, we have%
\begin{align}
\left\langle G\left(  z,T_{1},T_{2}\right)  \right\rangle  &  =\left\langle
S_{1}\left(  z,T_{1}\right)  S_{2}\left(  z,T_{2}\right)  \right\rangle
\nonumber\\
&  +\min\left\langle F_{1}\left(  z,T_{1}\right)  F_{2}\left(  z,T_{2}\right)
\right\rangle \\
&  +\min\left\langle D_{1}\left(  z,T_{1}\right)  D_{2}\left(  z,T_{2}\right)
\right\rangle ,\nonumber
\end{align}
where min$\left\langle \cdot\right\rangle $ results in min$\left\langle
\cdot\right\rangle \ll\left\langle \cdot\right\rangle $. This result requires
two conditions: (i) The measurement interval should be longer than the
coherence time of the light field, i.e., $\Delta T>\Delta t$, and (ii) the
rain removal effect of this method is inversely proportional to the
integration time of a single measurement event. \begin{figure}[ptbh]
\centering
\fbox{\includegraphics[width=1\linewidth]{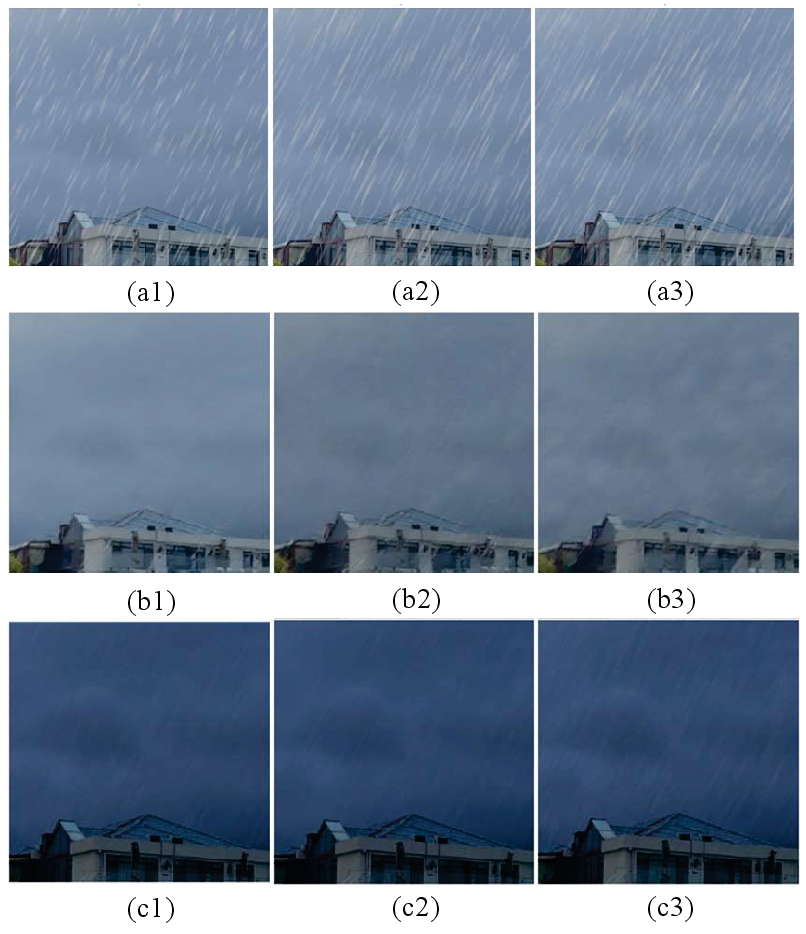}}\caption{Visual comparison
of the image rain removal results of different methods. Top row: Images with
different rain streak lengths (different integration times). The integration
times are (a1): 20 ms, (a2): 50 ms, and (a3): 80 ms. The results in middle row
are produced by the Pre-Net method, while the results in bottom row are
produced by our method. Each rain removal image is reconstructed from 30
measurements. }%
\label{fig:false-color}%
\end{figure}\begin{figure}[ptbh]
\centering
\fbox{\includegraphics[width=1\linewidth]{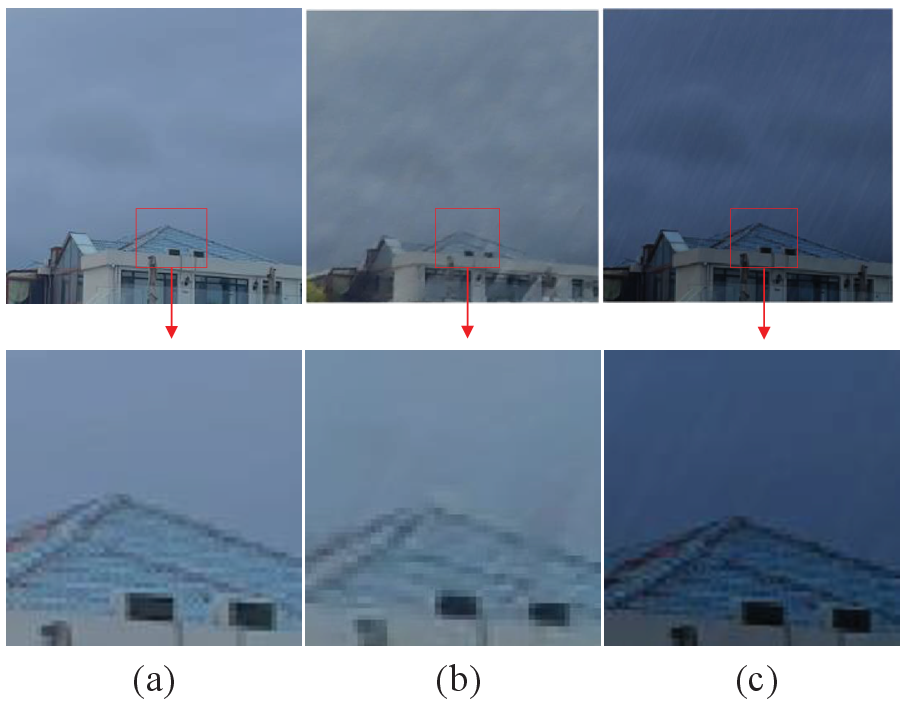}}\caption{Visual comparison
of the details of idle image (a), rain-removed image obtained by the PRe-Net
method (b), and the image reconstructed with our method (c). }%
\label{fig:false-color}%
\end{figure}\begin{figure}[ptbh]
\centering
\fbox{\includegraphics[width=1\linewidth]{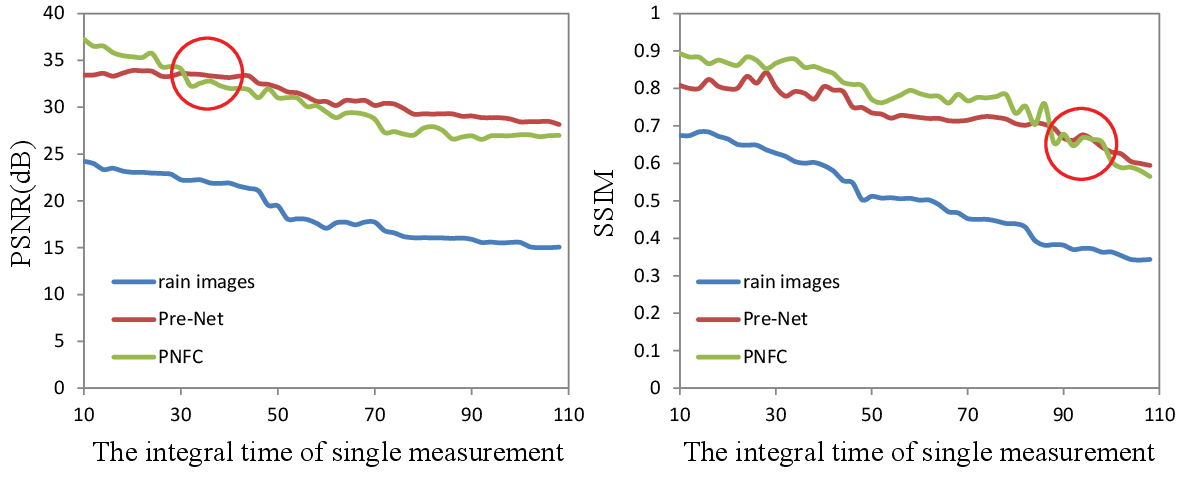}}\caption{PSNR and SSIM
curves of rain images and reconstructed images via our method (PNFC) and
images reconstructed via the conventional method (Pre-Net). The red circle
represents the inflection point of the two methods.}%
\label{fig:false-color}%
\end{figure}

A set of simulation and experimental results are used to present the rain
removal effect of our method. Figure 4 compares the rain images, the
reconstructed rain-removed images obtained by the traditional deep learning
method (Pre-Net) [30], and the reconstructed rain-removed images obtained by
our method. Obviously, in the images reconstructed by measuring the
second-order photon number fluctuation correlation in the time domain, the
influence of rain is almost completely eliminated by limiting the number of
measurements. The details of the image are clearer than those of the image
reconstructed using the traditional method (Fig. 5). However, the brightness
of the reconstructed image is slightly darker than the image without rain
because the fluctuating photons are eliminated through photon number
correlation measurements. The quantitative results (Fig. 6) show that the
effectiveness of this method is inversely proportional to the integration
time. The deraining quality of this method is better than that of traditional
deep learning methods when the integration time of the measurement events is
short. Figure 7 shows the actual effect of removing rain. \begin{figure}[ptbh]
\centering
\fbox{\includegraphics[width=1\linewidth]{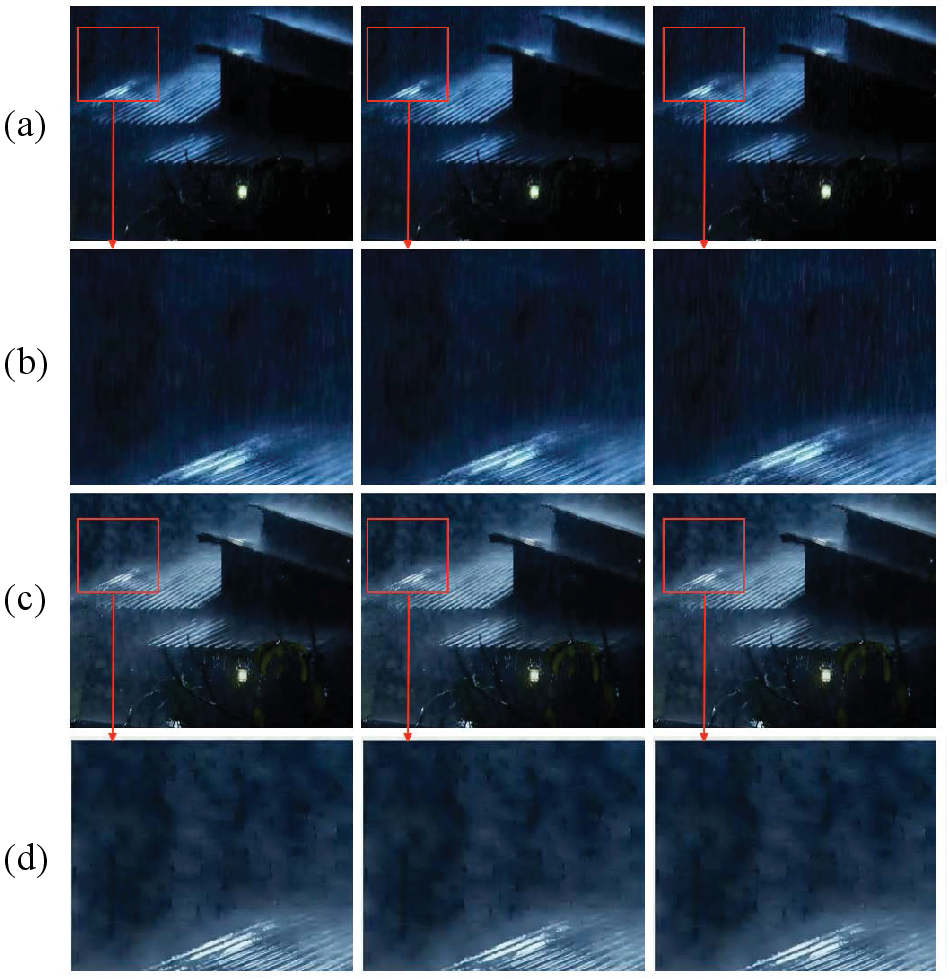}}\caption{Visual comparison
results of images with the rain removed through the second-order photon number
fluctuation correlation (a) and its details (b). The rain-removed images
obtained by the Pre-Net method (c) and its details (d). }%
\label{fig:false-color}%
\end{figure}

In conclusion, the statistical property and temporal second-order correlation of 
photons interacting with falling raindrops is investigated. We find that falling 
raindrops cause photon number fluctuations in pixels. However, the rain streak 
photons and rain fog photons introduced by falling raindrops results in the absence 
of a stable temporal second-order photon number correlation, while the photons that 
do not interact with raindrops are opposite. The rain streaks and rain fog are 
eliminated by measuring temporal second-order photon number fluctuation correlations 
when certain conditions are met. Thus, clean images can be reconstructed. The 
effectiveness of this quantum image rain removal method is inversely proportional 
to the integration time of a single measurement event. When the integration time 
of a single measurement event is short, the rain removal effect of this method is 
superior to that of traditional deep learning methods. This quantum rain removal 
method does not rely on any dataset of network model, and it has broad adaptability.

This work was supported by the Natural Science Foundation of Shandong Province
(ZR2022MF249) and the National Natural Science Foundation (12274257).

\end{document}